\documentclass[journal,10pt,twocolumn]{IEEEtran}

\newtheorem{thm}{Theorem}
\newtheorem{lemma}{Lemma}
\IEEEoverridecommandlockouts
\usepackage{amsmath,amssymb,amsfonts,float,graphicx,steinmetz,bm,subfigure,multirow,url,color,xcolor,etoolbox,ifthen,cite}
\usepackage{mathtools}
\usepackage{textcomp}
\usepackage{tikz}

\DeclarePairedDelimiter\floor{\lfloor}{\rfloor}

\DeclareMathSizes{10}{9.5}{6}{5}

\definecolor{lime}{HTML}{A6CE39}
\DeclareRobustCommand{\orcidicon}{%
	\begin{tikzpicture}
	\draw[lime, fill=lime] (0,0) 
	circle [radius=0.16] 
	node[white] {{\fontfamily{qag}\selectfont \tiny ID}};
	\draw[white, fill=white] (-0.0625,0.095) 
	circle [radius=0.007];
	\end{tikzpicture}
	\hspace{-2mm}
}

\foreach \x in {A, ..., Z}{%
	\expandafter\xdef\csname orcid\x\endcsname{\noexpand\href{https://orcid.org/\csname orcidauthor\x\endcsname}{\noexpand\orcidicon}}
}

\makeatletter 

\begin{document}             

\markboth{This work has been submitted to the IEEE for possible publication.Copyright may be transferred without notice.}{}

\title{ \huge
Large Intelligent Surfaces with Channel Estimation Overhead: Achievable Rate and Optimal Configuration
\thanks{This work was supported by the General Research Fund of the Hong Kong Research Grants Council (grant number 16202918). N. K. Kundu was also supported by the Hong Kong PhD Fellowship Scheme (PF17-00157).}
\vspace{-0.1in}
}


\author{
Neel Kanth Kundu\orcidA{}, {\em Student Member, IEEE} and
Matthew R. McKay\orcidB{}, {\em Fellow, IEEE}
\thanks{
The authors are with the Department of
Electronic and Computer Engineering, The Hong Kong University of
Science and Technology, Clear Water Bay, Kowloon, Hong Kong
(e-mail: {\tt nkkundu@connect.ust.hk}, {\tt m.mckay@ust.hk}).}
\vspace{-0.4in}
}


\maketitle

\begin{abstract}

Large intelligent surfaces (LIS) present a promising new technology for enhancing the performance of wireless communication systems. {\color{black} Realizing the gains of LIS requires accurate channel knowledge, and in practice the channel estimation overhead can be large due to the passive nature of LIS. Here, we study the achievable rate of a LIS-assisted single-input single-output communication system, accounting for the pilot overhead of a least-squares channel estimator.} {\color{black} We demonstrate that there exists an optimal $K^{*}$, which maximizes achievable rate by balancing the power gains offered by LIS and the channel estimation overhead. We present analytical approximations for $K^{*}$, based on maximizing an analytical upper bound on average achievable rate that we derive, and study the dependencies of $K^*$ on statistical channel and system parameters.}


\end{abstract}


\IEEEpeerreviewmaketitle
\vspace{-0.2in}
\section{Introduction}
Large intelligent surfaces (LIS) are a new physical layer technology that may play an important role in 6G wireless systems \cite{basar2019wireless,renzoRISsurvey20}. 
LIS do not require active components for signal reception and transmission, and are energy efficient \cite{basar2019wireless,renzoRISsurvey20,huang2019reconfigurable}.
Numerous studies so far have focused on optimizing the configuration of the LIS elements, referred to as ``passive beamforming'', using different criteria (e.g., \cite{ruiwu2018intelligent,yang2019intelligent,huang2019reconfigurable}).



Some recent works have studied the performance of LIS-assisted communications \cite{basar2019wireless,kudathanthirige2020performance,han2019large,kundu2020risassisted}, under the assumption that perfect channel knowledge is available at the transmitter, and without taking into account the overhead of acquiring the channel knowledge. This assumption may be restrictive for LIS, since the required pilot duration needed for performing channel estimation usually scales with the number of LIS elements, due to their passive nature \cite{zheng2019intelligent,jensen2019optimal}. Since LIS systems are typically envisioned to contain a large number of elements, this overhead could potentially be a limiting factor. 

To study this issue, here we investigate the performance of a LIS-assisted {\color{black} single-input single-output (SISO) system, incorporating the pilot overhead incurred with a least-squares (LS) channel estimator.} {\color{black} We consider a configuration for which the LIS elements are sufficiently spaced such that the corresponding channels experience independent fading {\color{black}(see \cite{renzoRISsurvey20})}, assumed to be Nakagami-$m$ distributed. } Our study shows that as the number of LIS elements $K$ increases, initially the average achievable rate increases due to the increased power gain provided by the LIS \cite{basar2019wireless,kudathanthirige2020performance}. However, beyond a certain point $K^{*}$, the penalty due to channel estimation overhead dominates, and the achievable rate starts decreasing monotonically. This behavior is in contrast to the behavior observed when one ignores channel estimation overhead, for which the performance has been shown to increase monotonically with $K$ \cite{basar2019wireless,kudathanthirige2020performance,ruiwu2018intelligent}. {\color{black} By deriving an upper bound on average achievable rate, we present analytical approximations for the optimal number of LIS elements $K^*$ used for communications, which we show varies with the statistical properties of the channel and the system. }

{\color{black} The trade-off between achievable rate and $K$ that we observe is consistent with previous results, which were presented under different modelling assumptions \cite{zheng2019intelligent,yang2019intelligent}. Specifically, considering LIS configurations with tightly packed elements (such that closely spaced elements experience highly correlated fading), it was shown through simulations that a strategy designed to group neighboring LIS elements can  improve performance by lowering the channel estimation overhead. An optimal grouping ratio was also studied empirically. 

We additionally point out recent works \cite{zappone2020overhead,zappone2020optimal} that have accounted for channel estimation overhead when considering beamforming design and resource allocation for LIS-assisted communications. Of most direct relevance to our study is  \cite{zappone2020optimal}, which proposed an adaptive greedy algorithm for selecting the optimum number of LIS elements to activate in order to maximize the channel-specific energy or spectral efficiency. While accounting for an overhead factor, the algorithm assumed the availability of perfect channel state information. The contribution \cite{zappone2020optimal} differs from our current study, where we investigate the number of LIS elements $K^{*}$ that maximizes the average achievable rate, which we characterize analytically, and which depends only on statistical channel knowledge.



}



{ \color{black}

}

\section{System Model}
We consider a scenario where a single-antenna source (S) communicates with a single-antenna destination (D) with the help of a LIS. {\color{black} We assume that communication involves $K$ LIS elements. Here, $K$ may be interpreted as the total number of LIS elements available, or a selected subset of `activated' LIS elements (which does not vary with the instantaneous channel coefficients).} { \color{black}We assume that both the size of each element and the inter-element spacing are equal to half of the wavelength of the radio signal, such that the associated channels exhibit independent fading \cite{renzoRISsurvey20}.}
We consider a time-division duplex scheme, where first the direct D-S and cascaded D-LIS-S channel is estimated at S. Exploiting channel reciprocity, S then uses the channel estimate to control the LIS phase shifts using a physical back-haul link, prior to data transmission.
\subsection{Channel Estimation}
In the channel estimation phase, the received pilot signal at S during the $t$-th training step is given by {\color{black}
\begin{align}
    y_t &= \sqrt{P_{{\rm tr}}} ( \sqrt{\beta_d} h_d + \sqrt{\beta_l} \bm{h}^T {\rm diag}(\bm{\phi}_t)\bm{g}) x_t + n_t
    \label{e2.1}
\end{align}
}where $y_t \in {\mathbb C}$ is the received signal at S, $x_t \in \mathbb{C}$ with $|x_t|=1$ is the pilot symbol transmitted from D, $P_{{\rm tr}}$ is the pilot transmit power, $n_t  \sim {\mathcal CN} (0, \sigma_{\rm tr}^2)$ is the AWGN at S, and $\bm{\phi}_t  = [e^{j \theta_{t,1}}, \ldots, e^{j \theta_{t,K}} ]^{T} \in {\mathbb C}^{K}$ denotes the vector of phase shifts induced by the LIS, where $\theta_{t,k} \in [0, 2\pi]$. {\color{black} We assume lossless reflection for all active LIS elements. The distance dependent path-loss and large-scale fading of the direct D-S channel and the D-LIS-S cascaded channel is captured by  $ \beta_d$ and $\beta_l$ respectively.  
The vectors $\bm{h} = [ h_1, \ldots, h_K  ]^T \in {\mathbb C}^{K} $ and $\bm{g} = [ g_1, \ldots, g_K  ]^T \in{\mathbb C}^{K} $ represent the small-scale fading vectors for the S-LIS and D-LIS link respectively, whose elements are modelled as i.i.d Nakagami-$m$ random variables with power normalized to $1$. 
Moreover, the scalar 
$h_d \in{\mathbb C}$ reflects the small-scale fading coefficient for the direct S-D link, also assumed to be Nakagami-$m$ distributed with power normalized to $1$. Nakagami-$m$ is a general fading model which can capture a variety of fading environments ($m=1$ corresponds to Rayleigh fading and $m=\frac{(\kappa+1)^2}{2\kappa+1}$ corresponds to Rician fading, with $\kappa$ denoting the Rician factor \cite{goldsmith2005wireless}). We denote by $m_1$ and $m_2$ the Nakagami-$m$ fading parameters of the elements of $\bm{h}$ and $ \bm{g}$ respectively, and denote $m_3$ as the corresponding parameter of $h_d$.} In the following, we will define the training SNR as $ \gamma_{{\rm tr}} = P_{{\rm tr}}/\sigma_{\rm tr}^2$.


Due to the passive nature of the LIS elements, it is difficult to estimate $\bm{h}$ and $\bm{g}$ separately. Rather, the {\em cascaded} channel $\bm{h}^T {\rm diag}(\bm{g})= \bm{v}^T = [ v_1, \ldots, v_K  ] \in {\mathbb C}^{1 \times K} $ is estimated. With this definition, we can re-express (\ref{e2.1}) as
{\color{black}
\begin{equation}
    y_t = \sqrt{P_{{\rm tr}}} (\sqrt{\beta_d} h_d + \sqrt{\beta_l} \bm{v}^T \bm{\phi}_t) x_t + n_t \;.
    \label{e2.2}
\end{equation}
It is assumed that $\bm{h}$, $\bm{g}$, and $h_d$ remain constant during a channel coherence block of length $T_c$, and that this satisfies $T_p < T_c$, where $T_p$ is the pilot length. Collecting all the received signals $y_t$ from (\ref{e2.2}), and using the DFT-based design \cite{zheng2019intelligent,jensen2019optimal} for $\bm{\phi}_t$ across $t=1,2,\ldots, T_p $,  a linear measurement model is obtained \cite[eq.~9]{jensen2019optimal}. The least squares (LS) estimate of the direct and cascaded channel can then be obtained using \cite[eq.~10]{jensen2019optimal}, which exists when $T_p \geq K+1$. This necessary condition implies that the channel estimation overhead $T_p$ scales at least linearly with the number of LIS elements $K$ \cite{zheng2019intelligent,yang2019intelligent,jensen2019optimal}. The condition $T_p < T_c$ also implies that $K < T_c$, which we will assume throughout.
}
\subsection{Data Transmission}
From (\ref{e2.2}), and by channel reciprocity, the received signal at D during the data transmission phase can be expressed as {\color{black}
\begin{equation}
    y = \sqrt{P} (\sqrt{\beta_d} h_d+ \sqrt{\beta_l} \bm{\phi}^T \bm{v}) x + n
    \label{e2.6}
\end{equation}
}where $P$ is the data transmit power, $x$ is information symbol with ${\mathbb E}[|x|^2]=1$, and $n  \sim {\mathcal CN} (0, \sigma^2)$ is the AWGN at D. We denote $ \Bar{\gamma} = P/\sigma^2$ as the average transmit SNR during the data transmission phase. The vector $\bm{\phi} = [ \phi_1, \ldots, \phi_K  ]^T \in {\mathbb C}^{K}$ denotes the LIS phase shift vector during the data transmission phase which, as before, has unit modulus entries. 

For the case of $\bm{v}$ assumed known, the received SNR at D is maximized by setting {\color{black} \cite[eq.~7]{gao2020unsupervised}
\begin{equation}
    \phi_i = {\rm exp}\{j \phase{ h_d/v_i } \} , \; \; \; \forall \; i=1,2,\ldots K
    \label{e2.7} \,,
\end{equation}
where $ \phase{z}$ denotes the argument of complex number $z$. Practically, $\bm{v}$ and $h_d$ are unknown and are replaced with their estimated values. We consider the LS estimate $\hat{\bm{v}}_{{\rm ls}},  \hat{h}_{d_{\rm ls}} $ obtained from \cite[eq.~10]{jensen2019optimal}, giving
\begin{equation}
    \hat{\phi}_i = {\rm exp}\{j \phase{ \hat{h}_{d_{\rm ls}}/\hat{v}_{{\rm ls}_i} } \} , \; \; \; \forall \; i=1,2,\ldots K \; .
    \label{e2.8}
\end{equation}
}
\section{Overhead-Aware Achievable Rate}
We study the average achievable rate of the LIS system, accounting for channel estimation. This is given by \cite{massivemimobook} {\color{black}
\begin{equation}
    R = \left(1-\frac{T_p}{T_c} \right) {\mathbb E}\left[ \log_2\left(1+ \Bar{\gamma} | \sqrt{\beta_d} h_d + \sqrt{\beta_l} \hat{\bm{\phi}}^T \bm{v}  |^2 \right) \right]  \; .
    \label{e3.1}
\end{equation}
}
The effect of channel estimation overhead is captured by the pre-log factor, which increases with $T_p$. Consequently, this overhead also increases with $K$, since {\color{black} $T_p\geq K+1$ } is required for the existence of the LS channel estimate. 

\begin{figure*}[htp] 
\centering
\subfigure[$R$ versus $T_p$ ]{%
\includegraphics[width=0.5\textwidth]{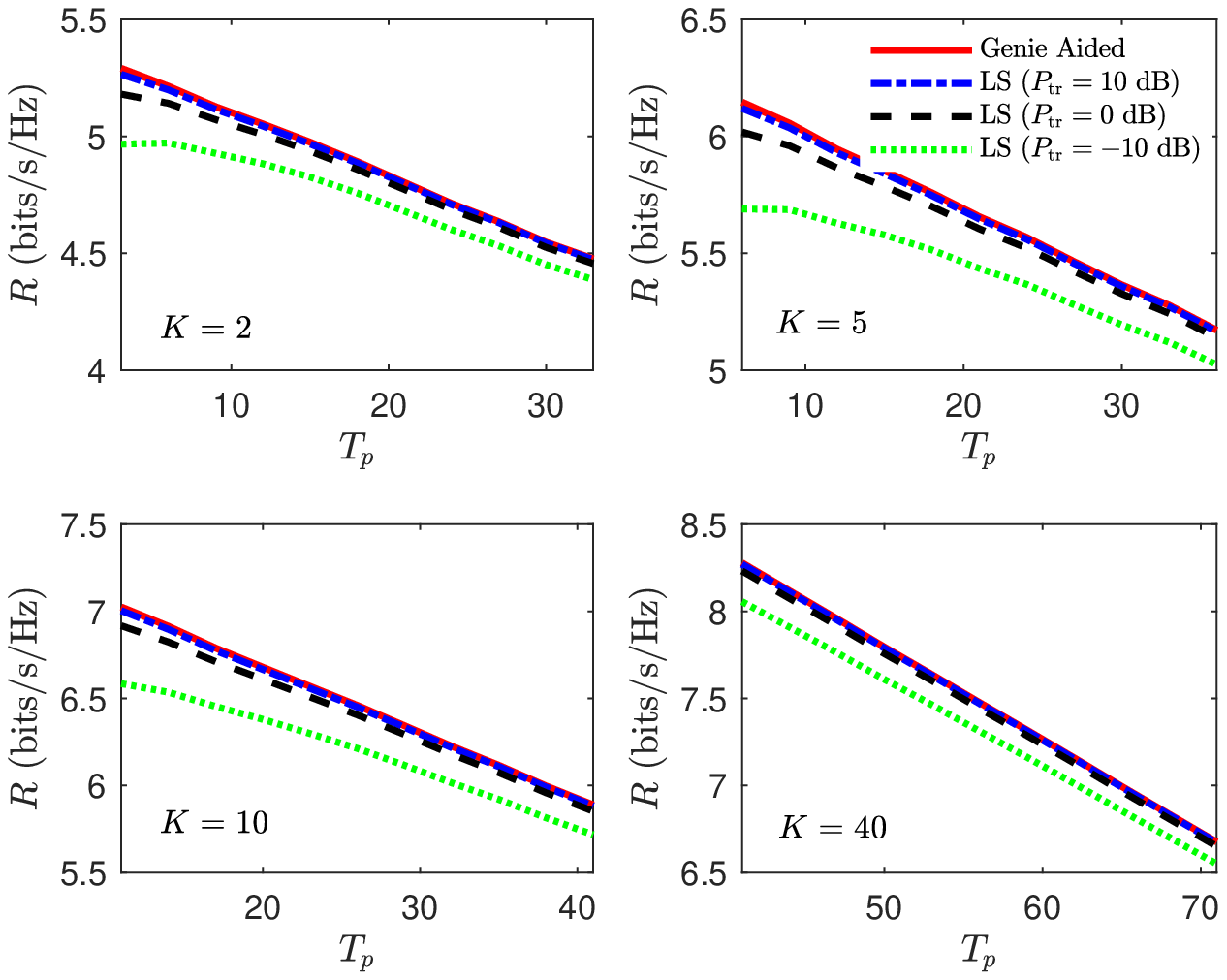}%
\label{fig3.1:a}%
}\hfil
\subfigure[Heatmap of $R\, (P_{\rm{tr}} = 0$ dB$)$ ]{%
\includegraphics[width=0.5\textwidth]{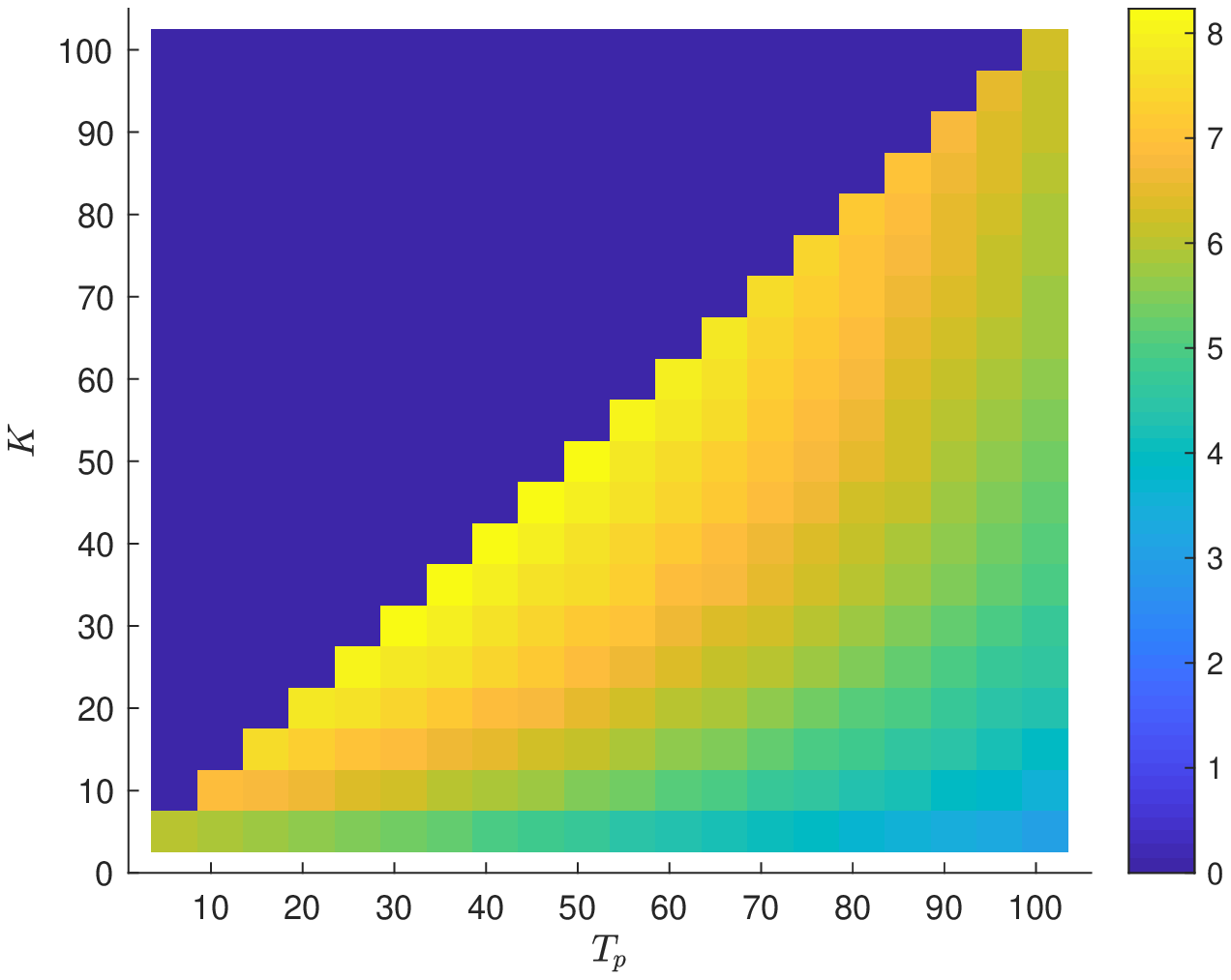}%
\label{fig3.1:b}%
}

\caption{{\color{black} Subplots in (a) show  achievable rate $R$ versus pilot length $T_p$ for different pilot power $P_{\rm tr}$ and different $K$. They also show a ``Genie Aided'' curve, which ignores channel estimation error and acts as a performance upper bound. Subfigure (b) shows a heat map of $R$ for different values of $K$ and $T_p \geq K +1$ with $P_{{\rm tr}} = 0$ dB. {\emph{General System Parameters (applicable to all):}}
$P= 0$ dB, $T_c= 196$ and Nakagami-$m$ fading parameter $m=0.5$ for all channels. The LIS is assumed to act like a scatterer, and the far field path loss model from \cite{di2020analytical} is applied. The distance between S and LIS is $d_1= 50$ m, between LIS and D is $d_2 = 5$ m, and between S and D is $d_3 = 60$ m. The path loss exponent for the direct channel is set as $3.5$, whereas for the cascaded channel it is set as $2$, and the reference path loss at a distance of $1$ m is set to $-30$ dB, and the noise variance is set to $\sigma^{2}= -80$ dBm \cite{ruiwu2018intelligent}. 
} }

\label{fig3.1}
\end{figure*}

{\color{black}
We start with a simulation example to demonstrate the joint effect of $T_p$ and $K$ on achievable rate. This is presented in Fig.\ \ref{fig3.1}, where we assume simulation parameters similar to \cite{ruiwu2018intelligent} (see figure caption for details). Numerical computation of (\ref{e3.1}) as a function of $T_p$ is shown in Fig. \ref{fig3.1:a} for different values of $P_{\rm tr}$ and $K$. For reference, it also shows ``Genie Aided'' curves, for which $\hat{\bm{\phi}}$ is replaced by $\bm{\phi}$ in (\ref{e3.1}) (i.e., channel estimation errors are ignored), which serves as a performance upper bound. A main observation is that, with the exception of scenarios for which both the number of LIS elements $K$ is very small and the pilot power $P{\rm tr}$ is very low, it is optimal to set the pilot length $T_p$ at its smallest possible value; i.e., $T_p = K + 1$.  
The heatmap of $R$, expressed as a function of $T_p$ and $K$ in Fig. \ref{fig3.1:b}, shows additionally that the rate observed along the main diagonal (i.e., for which $T_p=K+1$) first increases with $K$, before reaching a maximum, and then monotonically decreasing.  This empirically observed trade-off will be further explored in the following.

As a basis of our analysis, we first derive an analytical upper bound for $R$.  We apply a bounding approach since an exact computation of $R$ is difficult, due to the complexity of exactly describing the statistical distribution of the passive beamformer 
$\hat{\bm{\phi}}$ in (\ref{e3.2}), which depends on the residual angles arising due to channel estimation errors \cite{badiu2019communication}. We will show that the bound we derive is tight, and that it serves as a useful proxy for studying the trade-off of $R$ with respect to $K$.
}

\begin{lemma}
The achievable rate $R$ is upper bounded by  {\color{black}
\begin{equation}
  \tilde{R} = \left(1-\frac{T_p}{T_c} \right) \log_2(1+\Bar{\gamma}(aK^2+bK+c))  
  \label{lemm1}
\end{equation}
where
\begin{align}
   a =  \beta_l \delta_{1}^2 \delta_{2}^2, \;
   b =  \beta_l (1-\delta_{1}^2 \delta_{2}^2)  + 2 \sqrt{\beta_d \beta_l} \delta_{1} \delta_{2} \delta_{3}, \; c = \beta_d \;,
   \label{e3.1a}
\end{align}
with $\delta_1 = {\mathbb E}[|h_i|], \; \delta_2 = {\mathbb E}[|g_i|], \delta_3 = {\mathbb E}[|h_d|]$, given by
\begin{equation}
    \delta_l = \frac{ \Gamma \left( \frac{2m_{l}+1}{2}\right)}{ \sqrt{m_l} \Gamma(m_{l})} \;, \; \; {\rm for} \; \;  l=1,2,3 \;.
    \label{e3.1b}
\end{equation}
}
\label{l1}
\end{lemma}
{\color{black}
\begin{IEEEproof}
Our proof involves applying two successive upper bounds.  First, we upper bound (\ref{e3.1}) by  applying Jensen's inequality, which gives  
\begin{equation}
    R \leq \left(1-\frac{T_p}{T_c} \right)  \log_2\left(1+ \Bar{\gamma} {\mathbb E} \left[ | \sqrt{\beta_d} h_d + \sqrt{\beta_l} \hat{\bm{\phi}}^T \bm{v}  |^2 \right] \right)  \; .
    \label{e3.2}
\end{equation}
Next, this is further upper bounded by replacing $\hat{\bm{\phi}}$ with $\bm{\phi}$; i.e., ignoring estimation error, and assuming that perfect knowledge of $\bm{v}$ is used to design $\bm{\phi}$. Substituting $\bm{\phi}$ from (\ref{e2.7}), we obtain
\begin{align}
{\mathbb E} \left[ |\sqrt{\beta_d} h_d + \sqrt{\beta_l} \bm{\phi}^T \bm{v} |^2 \right] &=\beta_d \underbrace{{\mathbb E} \left[ |h_d|^2 \right]}_{z_1}  + \beta_l \underbrace{{\mathbb E} \left[ \left(\sum_{i=1}^{K} |h_i||g_i| \right)^2 \right]}_{z_2} \nonumber \\
&+ \sqrt{\beta_d \beta_l} \underbrace{2{\mathbb E} \left[ \left(\sum_{i=1}^{K} |h_i||g_i| |h_d| \right) \right]}_{z_3} \; .
 \label{e3.3}
\end{align}
Now, we have $z_1 =1$, and for evaluating $z_2$ and $z_3$ we need the following result for a Nakagami-$m$ random variable $w$ (with ${\mathbb E}[w^2] =1$):  
\begin{align}
  {\mathbb E} \left[ w \right] & = \int_{0}^{\infty} \frac{2m^m w^{2m}}{\Gamma(m)} e^{ -mw^2 } dw \stackrel{(a)}{=} \frac{ \Gamma \left( \frac{2m+1}{2}\right)}{\Gamma(m)  \sqrt{m}} \;,
  \label{e3.3b}
\end{align}
where $(a)$ follows from \cite[eq.~3.326(2)]{gradshteyn}. 
Using (\ref{e3.3b}), $z_2$ can be evaluated as
\begin{align}
    z_2  & = \sum_{i=1}^{K}{\mathbb E}\left[|h_i|^2 |g_i|^2 \right] + \sum_{i=1}^{K}{\mathbb E}\left[ |h_i| |g_i| \right] \left( \sum_{\substack{ j=1 \\ j \neq i }}^{K}{\mathbb E}\left[|h_j| |g_j| \right] \right) \nonumber \\
    &=  K  + K(K-1) \delta_{1}^{2} \delta_{2}^{2}  \;,
    \label{e3.3c}
\end{align}
where we use the independence of the two channels $\bm{h}$ and $\bm{g}$. Similarly, $z_3$ is evaluated as $z_3 = 2K \delta_{1}\delta_{2}\delta_{3} $.
Finally, combining $ z_1, z_2$ and $z_3$ we obtain the result in (\ref{lemm1}). 
\end{IEEEproof}
}


%
%

\begin{figure}[htp] 
\centering
\includegraphics[width=0.4\textwidth]{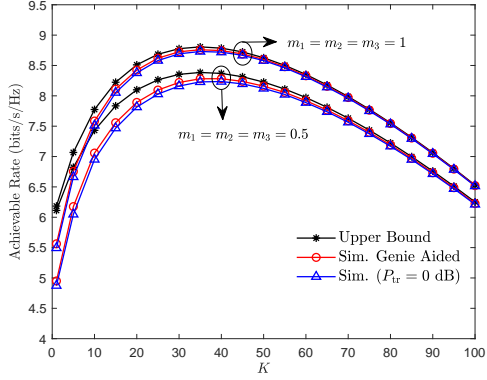}%
\caption{{\color{black} Comparison of the achievable rate upper bound of Lemma \ref{l1} and the exact achievable rate (\ref{e3.1}), with $P_{{\rm tr}}= 0$ dB. For further comparison, a ``Genie Aided'' scenario is also simulated, corresponding to (\ref{e3.1}) but with $\hat{\bm{\phi}}$ replaced by $\bm{\phi}$. The pilot duration is chosen as $T_p =K+1$, and the other system parameters are the same as the \emph{General System Parameters} defined in the caption of Fig. \ref{fig3.1}, except for the Nakagami-$m$ fading parameters which are indicated on the plot.
}
}
\label{fig4.2}
\end{figure}

Note that when viewed as a function of $T_p$, the bound (\ref{lemm1}) is maximized by choosing its minimum value, {\color{black} i.e., $T_p = K+1$, which aligns with the empirical observations from Fig.\ \ref{fig3.1:a} for practically reasonable $K$ values. } The accuracy of the bound is confirmed in Fig. \ref{fig4.2}, where it is compared with the exact achievable rate (\ref{e3.1}), computed numerically, assuming a pilot power of $P_{{\rm tr}}= 0$ dB. In addition, a ``Genie Aided'' curve is again provided for reference. {\color{black} Results are shown as a function of $K$, with $T_p=K+1$, and for two Nakagami-$m$ fading parameters. It is observed that, for both fading scenarios, the upper bound (\ref{lemm1}) from Lemma \ref{l1} matches very closely with the Genie Aided curve, and agrees well with the exact achievable rate (\ref{e3.1}), despite the fact that $P_{{\rm tr}}$ is quite low. Importantly, the upper bound captures the same general trend as the simulated curves, including the approximate location of the rate-maximizing $K$. 
} 

\section{Optimal LIS Configuration}
Based on the result of Lemma \ref{l1}, we look for an analytical approximation for the rate-maximizing $K$. {\color{black} To this end, setting $T_p =K+1$, we recall that $T_c > T_p$ (by assumption), and hence $K+1 < T_c$. Therefore, we seek to solve
\begin{equation}
    K^{*} = \underset{0< K \leq T_c -1}{ {\rm arg \; max}} \;  \tilde{R}(K)
    \label{e3.4} \; ,
\end{equation}
with $\tilde{R}(K)$ given by
\begin{equation}
     \tilde{R}(K) = \left(1-\frac{K+1}{T_c} \right) \log_2\left(1+\Bar{\gamma}(aK^2+bK+c)\right) \;.
     \label{e15}
\end{equation}
Our solution is given in the following:}
\begin{thm} \label{theo1}
There exists a unique optimal solution $K^*$ to (\ref{e3.4}) obtained by numerically solving {\color{black}
\begin{equation}
    \frac{\Bar{\gamma}(T_c-K-1)(2aK+b)}{ 1+\Bar{\gamma}(aK^2+ bK + c)} = \ln(1+\Bar{\gamma}(aK^2+ bK+c)) 
    \label{e3.6}
\end{equation}}for $K$, and then rounding the solution to the nearest integer. 
\end{thm}

\begin{IEEEproof}
Computation of the first derivative of $\tilde{R}(K)$ yields: {\color{black}
\begin{align}
     \tilde{R}'(K & )= \frac{\Bar{\gamma}(T_c-K-1)(2aK+b)}{\ln(2) T_c(1+\Bar{\gamma}(aK^2+bK+c))} \nonumber \\ 
     & \hspace{2cm} - \frac{\log_2(1+\Bar{\gamma}(aK^2+bK+c))}{T_c} 
      \label{e3.5} \; \;  .
\end{align} }Since $a,b>0$, after some straightforward algebraic manipulations, it can be shown that $\tilde{R}''(K)<0 $, which implies that $\tilde{R}'(K) $ is a strictly monotonically decreasing function of $K$. Moreover, from (\ref{e3.5}) we find that $ \tilde{R}'(0) > 0$ and {\color{black}$ \tilde{R}'(T_c -1) < 0$.} This implies that $\tilde{R}(K)$ first increases, reaches a unique maximum, and then decreases as the argument $K$ increases from $0$ to {\color{black}$T_c -1$.} Thus, for {\color{black}$0< K \leq T_c -1$,} $\tilde{R}(K)$ is a concave function of $K$, and the unique optimal solution to (\ref{e3.4}) is obtained by solving $ \tilde{R}'(K) = 0$.
\end{IEEEproof}

While a closed-form solution to (\ref{e3.4}) is difficult in general, it can be easily solved numerically. Moreover, a closed-form solution is attainable at high SNR, as shown by the following:
\begin{thm} \label{theo2}
As $\Bar{\gamma} \to \infty$ (i.e., at high transmit SNR), eq.\ (\ref{e3.4}) admits the solution 
{\color{black}
\begin{equation}
    K^{*} = \floor*{ \frac{T_c-1}{W\left(e\sqrt{\Bar{\gamma}a}(T_c -1)\right)}  + \frac{1}{2} }\;,
    \label{e3.7}
\end{equation}
}where $W(x)$ denotes Lambert's W-function \cite{corless1996lambertw}, $e$ denotes Euler's number, and $ \floor*{.} $ is the floor operation.
\end{thm}
\begin{IEEEproof}
{\color{black}
As $\Bar{\gamma} \to \infty$, it is easy to show that $\log_2 (1 + \psi \bar{\gamma} ) = \log_2 ( \psi \bar{\gamma} ) + o(1)$, for some constant $\psi$. Hence, $\tilde{R}(K)$ in (\ref{e15}) can be expressed as
\begin{equation}
    \tilde{R}_1(K)= \left(1-\frac{K+1}{T_c} \right)\log_2(\Bar{\gamma} a K^2) + o(1) \; .
    \label{e3.8}
\end{equation}
Taking the leading order term, and setting the first derivative of (\ref{e3.8}) w.r.t $K$ equal to 0, we obtain \begin{equation}
    \frac{T_c-1}{K} = 1 + \ln( \sqrt{\Bar{\gamma}a} K) \;.
    \label{e3.9}
\end{equation}}Letting $ t= \ln( \sqrt{\Bar{\gamma}a} K)$, after some algebra, (\ref{e3.9}) can be equivalently expressed as {\color{black}
\begin{equation}
    (1+t)e^{(1+t)}= e\sqrt{\Bar{\gamma}a}(T_c-1)
    \label{e3.10} \; .
\end{equation}}Using the definition of Lambert's W function \cite{corless1996lambertw}, this can be solved for $t$ as 
{\color{black}
\begin{equation}
    t=W\left(e\sqrt{\Bar{\gamma}a}(T_c-1)\right)-1 \;.
    \label{e3.12}
\end{equation}}After some further algebraic manipulations, including using the property $x=e^{W(x)}W(x)$ \cite{corless1996lambertw} and rounding off the solution to the nearest integer, we obtain the result in (\ref{e3.7}).

\end{IEEEproof}

\begin{figure}[htp] 
\centering
\subfigure[$K^*$ versus $T_c$ ]{%
\includegraphics[width=0.5\linewidth]{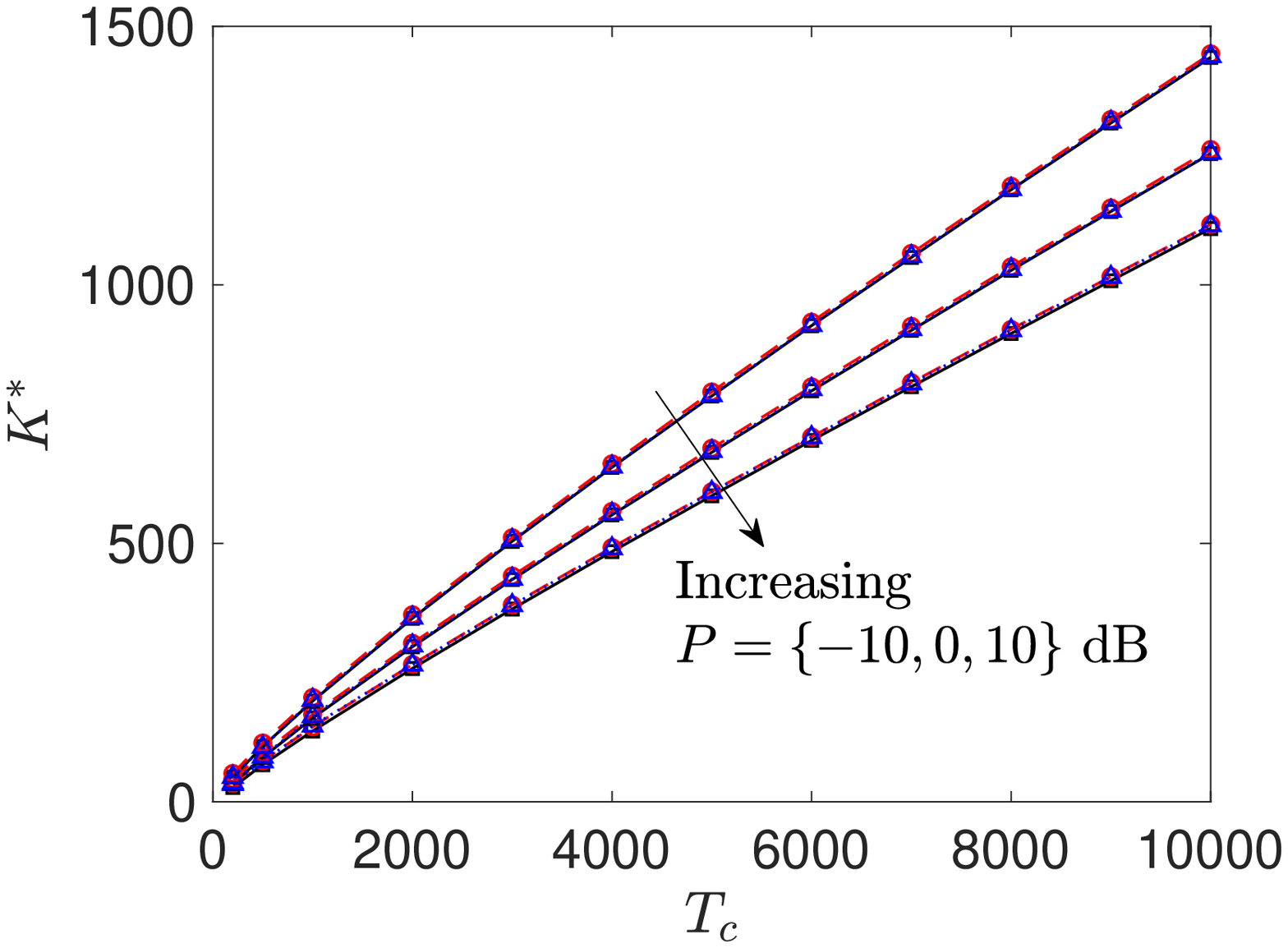}%
\label{fig4.3:a}%
}\hfil
\subfigure[$K^*$ versus $P$ ($T_c=2000$)  ]{%
\includegraphics[width=0.5\linewidth]{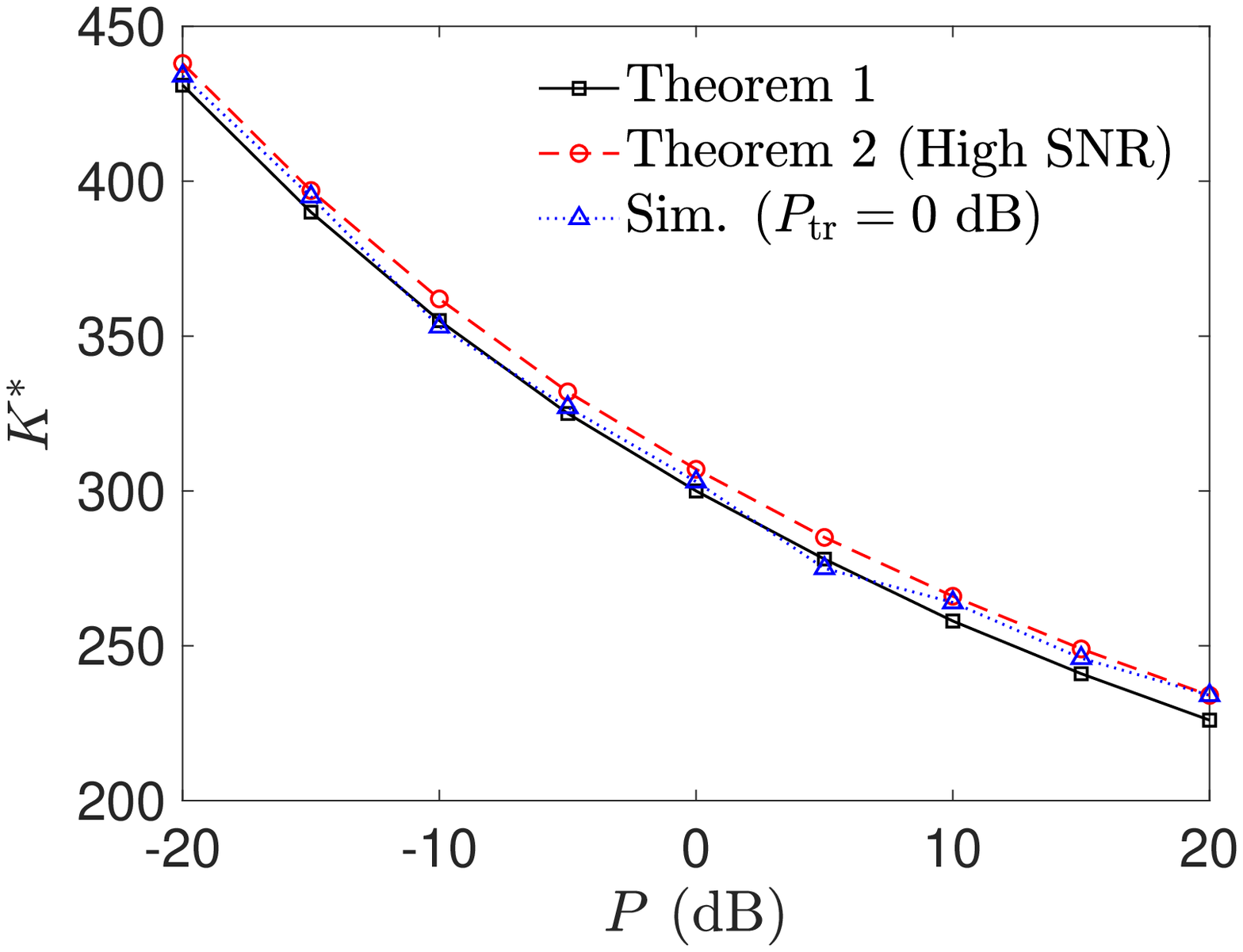}%
\label{fig4.3:b}%
}

\caption{{\color{black} The plots show the optimized number of LIS elements $K^{*}$ as a function of (a) the coherence time $T_c$, and (b) the transmit power $P$. Results are shown based on the analytical formulas in Theorems \ref{theo1}, \ref{theo2}, and numerical optimization of the exact achievable rate function (\ref{e3.1}) at $P_{{\rm tr}}=0$ dB. The system parameters are the same as the \emph{General System Parameters} defined in the caption of Fig. \ref{fig3.1}, except for $P$ and $T_c$ which are indicated on the plots.
} 
}
\label{fig4.3}
\end{figure}

{\color{black} 
It is seen from (\ref{e3.6}) and (\ref{e3.7}) that the optimized number of LIS elements $K^*$ depends on statistical channel parameters, including the coherence block length $T_c$ and the average transmit SNR $\bar{\gamma} = P / \sigma^2$.  The general behaviour, as a function of these statistical parameters, is shown in Figs. \ref{fig4.3:a} and \ref{fig4.3:b}. The figures compare $K^{*}$ obtained from Theorems \ref{theo1} and \ref{theo2}, and the value $K^*$ obtained by numerically optimizing the exact achievable rate function (\ref{e3.1}) with respect to $K$ at $P_{{\rm tr}}=0$ dB. Observe that the analytical solutions are quite accurate.

The plots show that the optimal number of LIS elements, $K^*$, grows near-linearly with $T_c$, while varying inversely with $P$. This trend can be established more precisely, based on (\ref{e3.7}). Using the large-$x$ expansion for the Lambert's W-function, $W(x) = \ln{x}-\ln{\ln{x}} + O( \ln{\ln{x}} / \ln{x})$ \cite[eq.~(4.11)]{corless1996lambertw}, it follows that as $T_c \to \infty$, $K^*$ in (\ref{e3.7}) scales as $O( T_c / \ln T_c )$. Moreover, as $P \to \infty$, or equivalently $\bar{\gamma} \to \infty$, $K^*$ scales as $O( 1 / \ln \bar{\gamma} )$. It is important to note that with these scaling behaviors, the optimal choice of LIS elements $K^*$ can vary widely between applications with different statistical properties.  The value of $T_c$, for example, depends on user mobility (see \cite{massivemimobook} for a discussion of $T_c$ for different application scenarios).  Generally speaking, for systems with shorter coherence block lengths, a smaller $K^*$ becomes beneficial, since there is less opportunity for data transmission (within a fixed channel period), and smaller $K^*$ has the benefit of reducing channel overhead by reducing the required pilot length. 

The plots in \ref{fig4.3:b} reveal that the result in Theorem \ref{theo2} is quite accurate for a wide range of $P$, despite being derived under a high SNR assumption.
The decrease of $K^{*}$ with increasing $P$ (or $\Bar{\gamma}$) is explained by the fact that as $P$ increases, the effective gain of the cascaded channel is improved. Hence, it becomes beneficial to have less LIS elements, since the reduction in training overhead outweighs the relative beamforming gains offered by having more LIS elements, which are less significant when the effective cascaded channel gain is high.




}

It is important to note that the general trend observed in Fig. \ref{fig4.3:b} does \emph{not} imply that the achievable rate degrades for higher $P$. In fact, it is quite the opposite. This is seen from Table \ref{tab1}, which reports the $K^*$ obtained from Theorem \ref{theo1}, along with the corresponding values of $\tilde{R}$, and $R$ obtained from numerical evaluation of (\ref{e3.1}), with $P_{{\rm tr}} =0$ dB and $T_c=2000$. Intuitively, as $P$ increases, the received SNR is larger due to a better effective channel, and a lower training overhead brought by a smaller value of $K^*$ leads to a higher achievable rate.

\begin{table}[htp]
\centering
\caption{{\color{black}Optimum $K^{*}$ from Theorem \ref{theo1} and the corresponding upper bound on achievable rate $\tilde{R}$ (b/s/Hz) from (\ref{lemm1}) and  $R$ (b/s/Hz) obtained from numerical evaluation of (\ref{e3.1}) at $P_{{\rm tr}} = 0$ dB. The other system parameters are the same as the General System Parameters defined in the caption of Fig. \ref{fig3.1}, except for $T_c=2000$. 
}
}
\begin{tabular}{|l|l|l|l|l|l|l|l|}
\hline
$P$ (dB)   & $-10$ & $-5$ & $0$ & $5$ & $10$ & $15$ & $20$  \\ \hline \hline
$K^{*}$ & $355$  &  $325$   & $300$   & $278$  & $258$ & $241$ & $226$     \\ \hline
$\tilde{R}$  &   $10.74$  & $12.12$   & $13.52$  & $14.94$  &  $16.38$ &  $17.83$ &  $19.29$\\ \hline
$R$ &   $10.71$  & $12.09$   & $13.5$  & $14.92$  &  $16.36$ &  $17.79$ &  $19.26$\\ \hline
\end{tabular}
\label{tab1}
\end{table}


\section{Conclusion}

Channel estimation overhead can pose a significant performance bottleneck for LIS systems.  { \color{black} Our results suggest that an approach to addressing this problem is to apply pragmatic selection of the number of LIS elements $K^*$ used for communications,  based on statistical parameters. 
%

Knowledge of $K^{*}$ can help at the system design phase by providing guidance on the maximum LIS size to deploy, using application-specific statistical knowledge. It could also guide adaptive protocol designs that seek to activate only a subset $K^{*}$ of the total available LIS elements based on statistical channel conditions. Due to its dependence on statistical knowledge only, adaptation at only a low rate would be required.

Our study assumed a LS channel estimator, which required the pilot length to be least as large as $K+1$. This is the primary source of the channel estimation overhead.} It is possible that under alternative channel assumptions, such as models involving sparsity, more efficient channel estimation methods may be employed. In such cases, the overhead may be less of an issue. It remains to be seen however, whether such methods can lead to monotonic growth in the achievable rate of LIS systems for increasing $K$ (as suggested by studies which ignore channel estimation overhead), or whether a trade-off, and optimal configuration $K^*$, still exists.

{ \color{black} Our analysis considered a scenario comprising a single-antenna transmitter and receiver, aided by one LIS. Extensions to more general models could be further explored, } {\color{black} such as incorporating multiple LIS, multiple users, and multiple antennas, as well as  multi-carrier systems (see e.g.,  \cite{zheng2020intelligent,zheng2020double})}. { \color{black}
These are worthwhile directions to pursue in future studies. }

\bibliographystyle{IEEEtran}
\bibliography{IEEEabrv,KUNDU_WCL2020-1817R1}
\end{document}